\documentclass[12pt,amsfonts,amsmath]{article}
\usepackage[utf8]{inputenc}
\usepackage[T1]{fontenc}
\usepackage{lmodern}
\usepackage{times}
\usepackage{amssymb}
\usepackage{amsmath}
\usepackage{amsthm}
\usepackage{graphicx}
\usepackage{tikz} 
\usepackage{color}

\newcommand{\R}{{\mathbb{R}}}

\newtheorem{Proposition}{Proposition}

\newtheorem{theorem}{Theorem}

\newcounter{exercice}

\title{Beyond the brain: towards a mathematical  modeling of emotions}
\author{Benjamin Ambrosio\\
benjamin.ambrosio@univ-lehavre.fr\\
Normandie Univ, UNIHAVRE\\
LMAH, FR-CNRS-3335, ISCN\\ 76600 Le Havre, France}
\date{September 1 2020}
\begin{document}
\maketitle
\abstract{Emotions are a central key for understanding human beings and of fundamental importance regarding their impact in human and animal behaviors. They have been for a long time a subject of study for various scholars including in particular philosophers and mystics \cite{BookAri-2000,BookCan2020,BookNum1998,Book-Plato-Republic,Bookkam-2017,BookDes2010,BookRumi2004,BookHafez2017}. In modern science, the emotional phenomenon has attracted for a few decades an increasing number of studies, notably  in the fields of Psychology, Psychiatry, Neuroscience and  Biochemistry \cite{Cell-2016,Ado-1994,Ado-1998,Damasio2013,Ekman2006,Fei-2011,Fei-2013,Gos-2007,Gu2019,LeDoux1995,LeDoux2017,Lin2013,Lin2012,BookSal-1997,Tra-2006}. However, since our perception of emotions is not, so far, directly detectable nor recordable by our measure instruments, Physics and Mathematics have not been  so far used academically to provide a precise description of the phenomenon of feeling an emotion. Relying upon the works of O. Elahi \cite{Dur-2011,Book-O-Elahi2012} and on the hypothesis that the human soul and its psyche may manifest in ourselves (in both conscious and unconscious manner) in an analog way as electromagnetic waves \cite{Book-B-Elahi-2019}, we propose here a few mathematical descriptions consistent  with the human personal experience, of the feeling and cognition of emotions. As far as we know, such a mathematical description has never been provided before. It allows a quantitative (intensity) and qualitative (nature of feelings/frequency) of the emotional phenomenon which provides a novel scientific approach of the nature of the mind, complementary to the on going research of physiological manifestation of emotions. We anticipate such an approach and the associated mathematical modeling to become an important tool  to describe    emotions and their subsequent behavior. In  complement of the modeling of oscillations and brain dynamics\cite{Amb-2020-3,Char2018,Per-2019}, it provides a fruitful direction of research with potentially broad and deep impacts in both applied mathematics, physics, cognitive and behavioral sciences.   
\section{Introduction}
Emotions, viewed as a drive phenomenon occurring in human soul, has for a long time been discussed by great names of ancient philosophy\cite{Book-Plato-Republic}. More recently, the Freudian model of the self, with notably its conscious and unconscious parts as well as the id as the source of libidinal and aggressive instincts\cite{BookFreud1990} provides also a functional representation allowing a rational explanation connecting description of emotions and behavioral comportment. In the works of O. Elahi\cite{Book-B-Elahi-2018,Book-B-Elahi-2019,Book-O-Elahi2012} the psyche is only a small part of the soul, and the specific human characteristics (such as creativity, specific human intelligence or sensibility to ethics and morality which distinguish human beings from animals) result from the intrinsic nature of the human soul rather than an exclusive social and cultural influence.  Of interest for the present work, is that in the model of O. Elahi, the effects of human soul and its psyche are compared to the effects of electromagnetic waves \cite{Book-B-Elahi-2019}. One has to point out that an electromagnetic radiation has two essential characteristics:
\begin{enumerate}
    \item its spectrum with possibly a main frequency
    \item its intensity
\end{enumerate}
This two characteristics fits well the description of an emotion. We can think that 
the frequency of an emotion provides its quality: for example the qualitative feeling of a sexual instinct is different from one of an aggressive instinct. From a mathematical point of view, one could therefore describe these two feelings by two different frequencies. One can also think about a psychic state as the summation of distinct single frequency wave components in the same way  that an electromagnetic radiation can be decomposed in its spectrum individual sinusoidal components. Secondly, the intensity of an emotion relates to the strength with which it manifests in ourselves and eventually drives us to a specific behavior. For example, an emotional putsch\cite{Book-B-Elahi-2018} of anger, which is a very intense pic of anger taking control over us, may lead to a non-appropriate behavior regarding ethics and social rules. In the next two paragraphs we provide two mathematical tools describing 1) an intense pulse of anger  and 2) its control.  

\section{The putsch of anger as a sequence of Gaussian functions converging towards the Dirac Distribution}    
\subsection{A typical experience of anger}
In a daily basis, anger can arise in different ways. Some situations might repeat again and again: analog stimuli from external environment will induce similar responses in our conscious psychic state. Let us describe a typical experience. Your kid, or other close relative of yours, does something that you find really wrong. Very quickly, some energy arises; first it is not really conscious, but rapidly you feel it in your conscious. And then, suddenly, it expresses itself in a very intense way. It is a pic of anger and you felt an energy pushing you to throw the object that you were handling in your hand! Many people have experienced this kind of precise emotions in their psyche. For this typical experience, a mathematical analogy appears to be relevant: the convergence of a sequence of Gaussians functions towards the Dirac distribution. 
\subsection{Sequence of Gaussian functions and the Dirac distribution. Application to Anger}
The Gaussian function  in $\R^2$ is given by the following expression:
\begin{equation}
\label{eq:GaussFunc}    
f_\sigma(x,y)=\dfrac{1}{\sigma\sqrt{2\pi}}\exp{\bigg(-\dfrac{x^2+y^2}{2\sigma^2}\bigg)}
\end{equation}
The Gaussian function appears in various domain in mathematics and has a wide spectrum of applications. In the form given here it specifically gives the density of probability of mean $0$ and standard deviation $\sigma$. Of interest here, is the following theorem, well known in mathematics, which expresses that as a sequence $\sigma_n$ converges towards $0$, the sequence of functions $f_{\sigma_n}$ given in \eqref{eq:GaussFunc} converges towards the Dirac distribution\cite{Boc-1952,BookLSchwartz-1966}.
The Dirac distribution $\delta$ is defined as a linear form on the space of smooth functions  $C^\infty$ with compact support, denoted by $\mathcal{D}(\R^2)$ such that for every $\varphi \in \mathcal{D}(\R^2)$
\[\delta(\varphi)=\varphi(0)\]
The following theorem holds:
\begin{theorem}
\[
\lim_{n \rightarrow +\infty }f_{\sigma_n}=\delta \mbox{ in the sense of distributions}\]
\end{theorem}
The point here is to interpret this theorem in terms of energy and further relate it to emotions. We need to first recall  that the integral of the Gaussian over $\R^2$, whatever the (positive) value of $\sigma$, is $1$. Viewing the Gaussian function as an energetic density,  an interpretation of this theorem, is that as $\sigma_n$ goes to $0$ the distribution of the energy narrows and at the limit the whole energy is concentrated in the single point $0$. This describes quite well some manifestations of the anger, such as the putsch of anger described above. When a person is calm, the energy of anger is distributed in the whole psyche (here represented by $\R^2$). As soon the stimulus occurs the energy tends to focus on our point of attention and eventually leads to an anger putsch. The anger putsch corresponding to the Dirac. A representation of the analogy is given in figure \ref{fig:Dirac-Anger}. Note that, relating to the electromagnetic waves, the energy considered here is a density energy, \textit{i.e.} the intensity of anger at each point of the psyche. The frequency, not described in this example, appears implicitly in the fact that the qualitative energy of the anger is specific.  As specific as the the color blue ranges in frequency 620-680 THz.

\begin{figure}
    \centering
    \includegraphics[width=4cm]{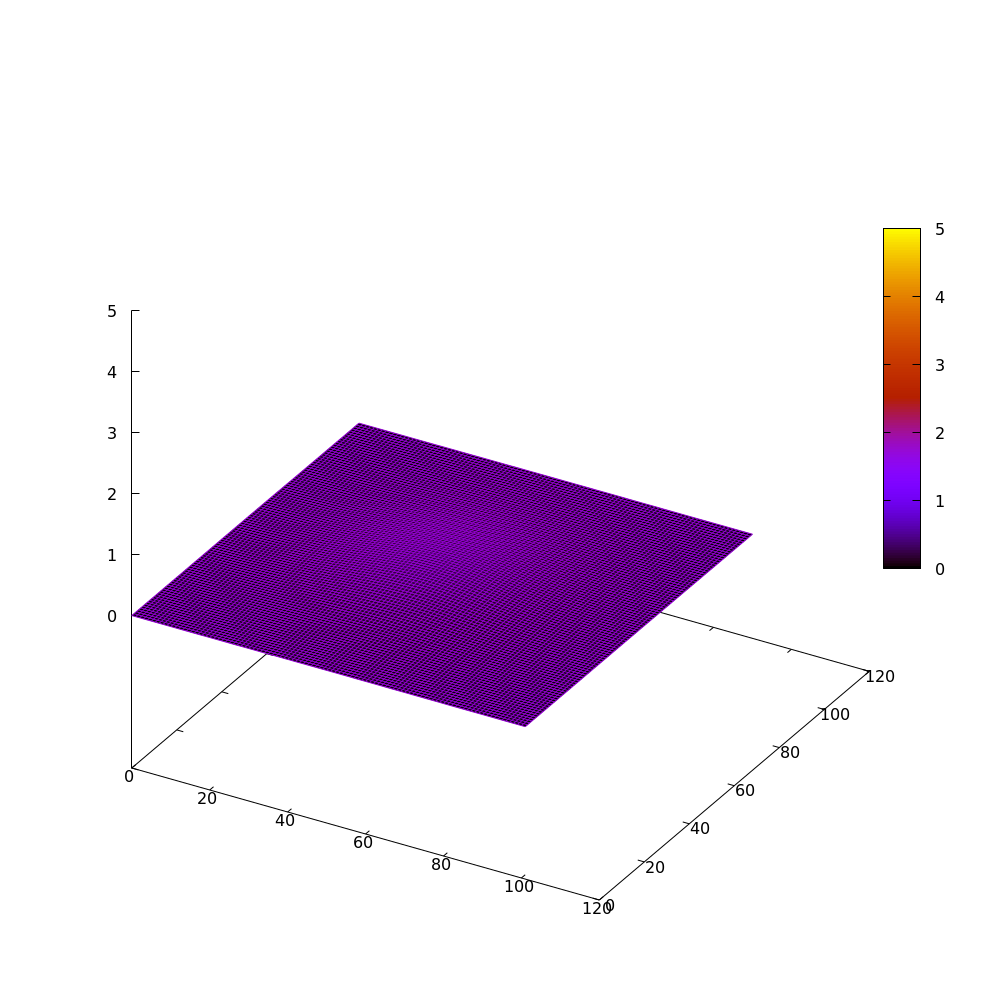}
       \includegraphics[width=4cm]{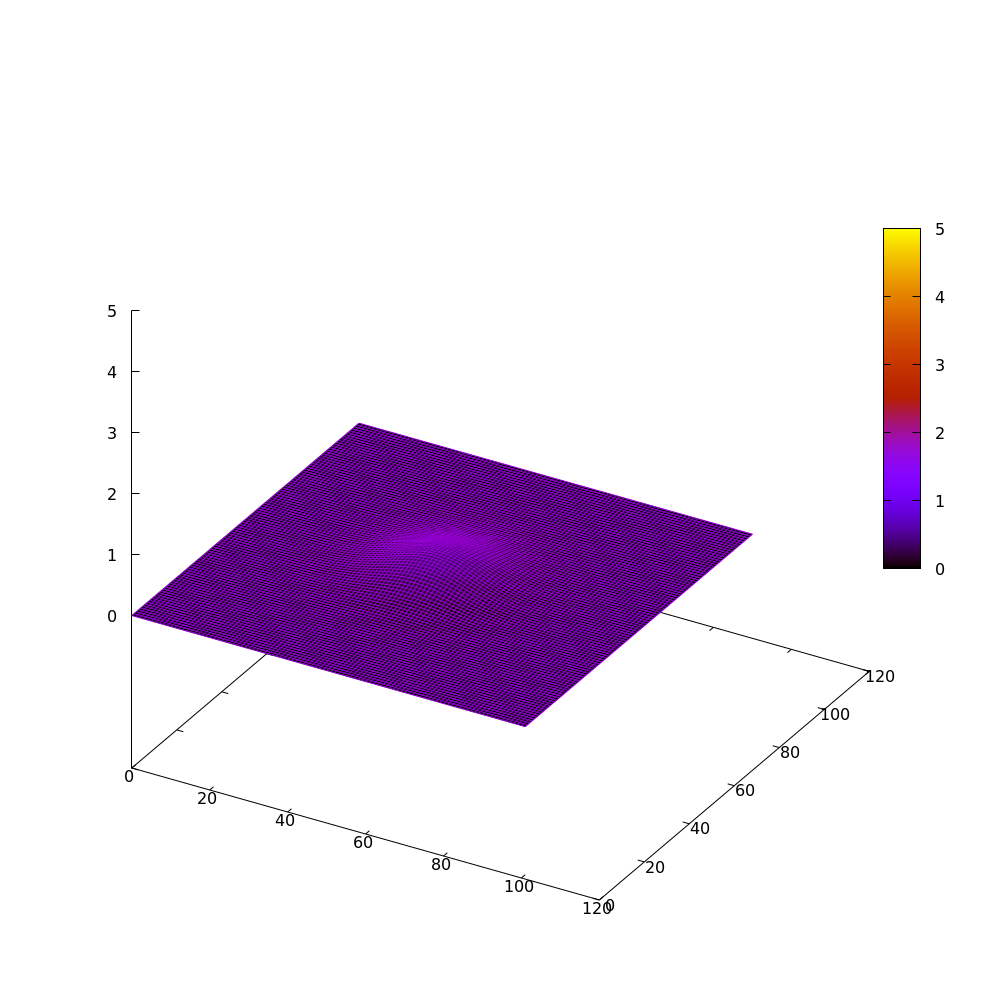}
          \includegraphics[width=4cm]{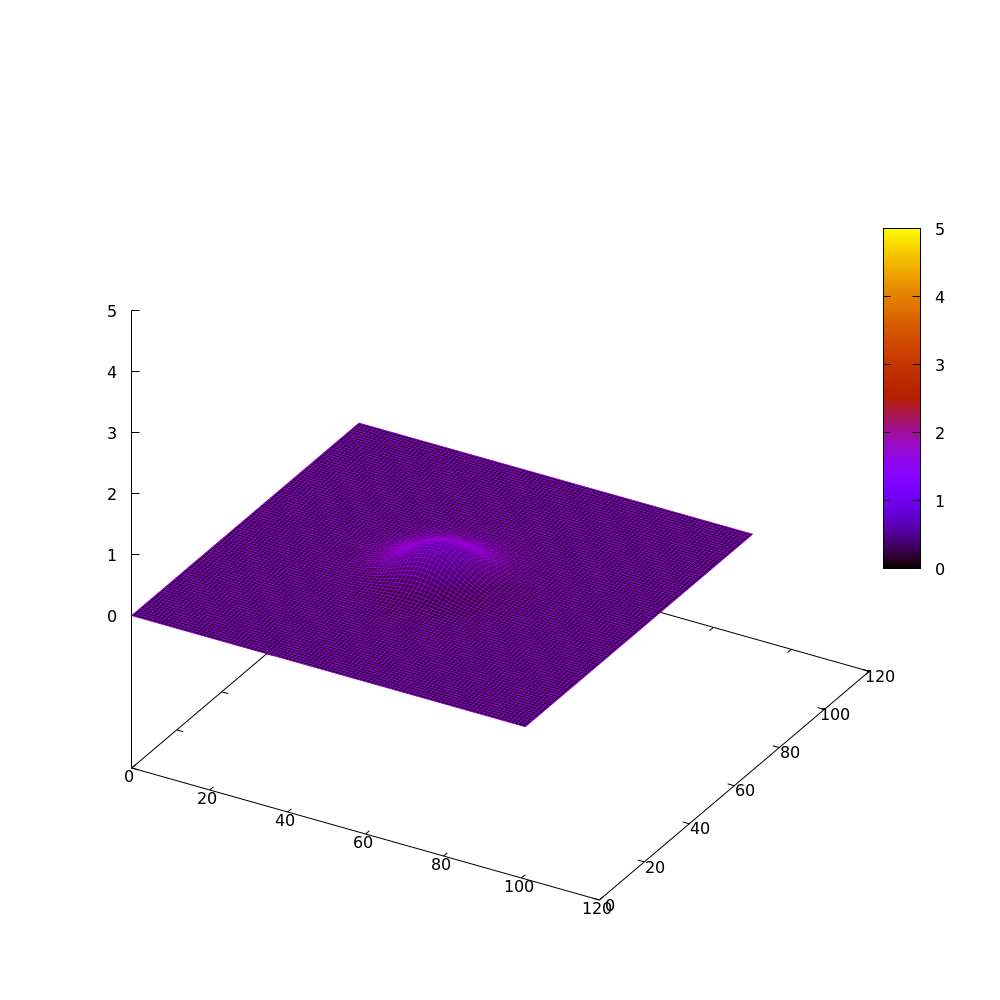}\\
             \includegraphics[width=4cm]{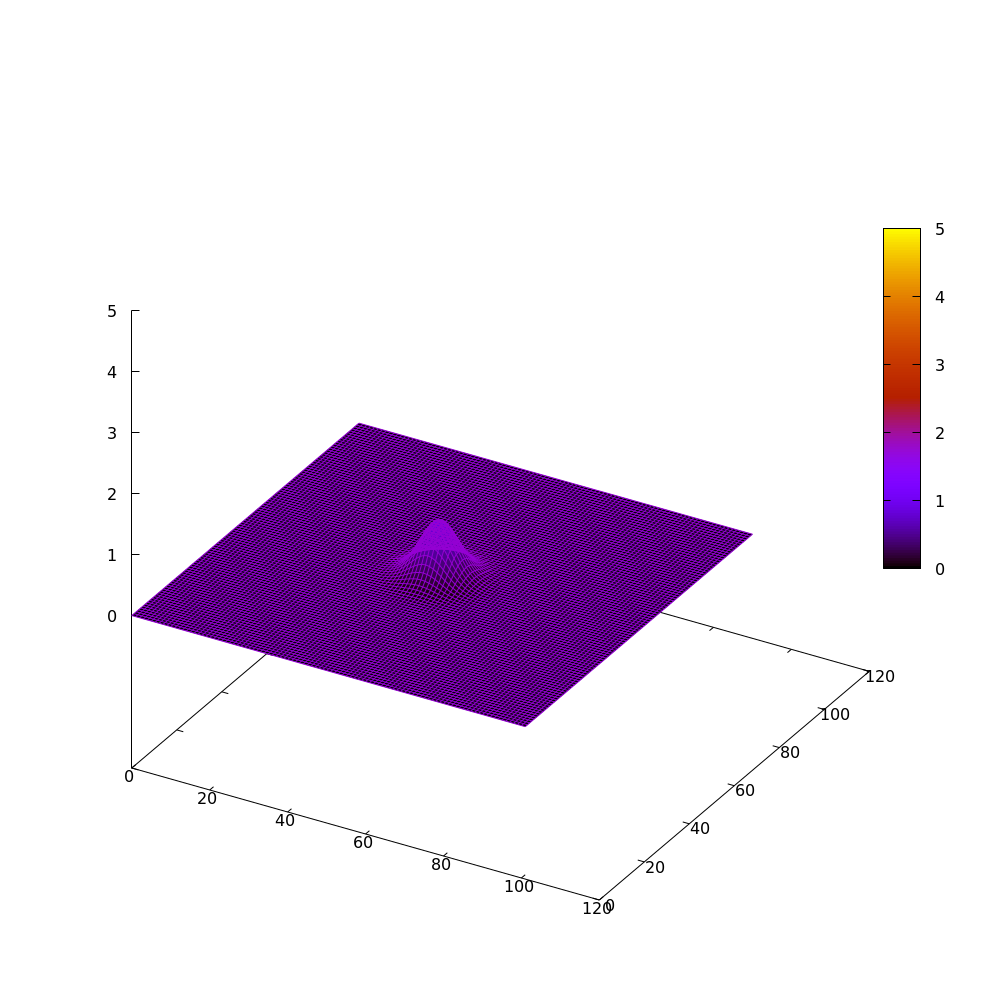}
                \includegraphics[width=4cm]{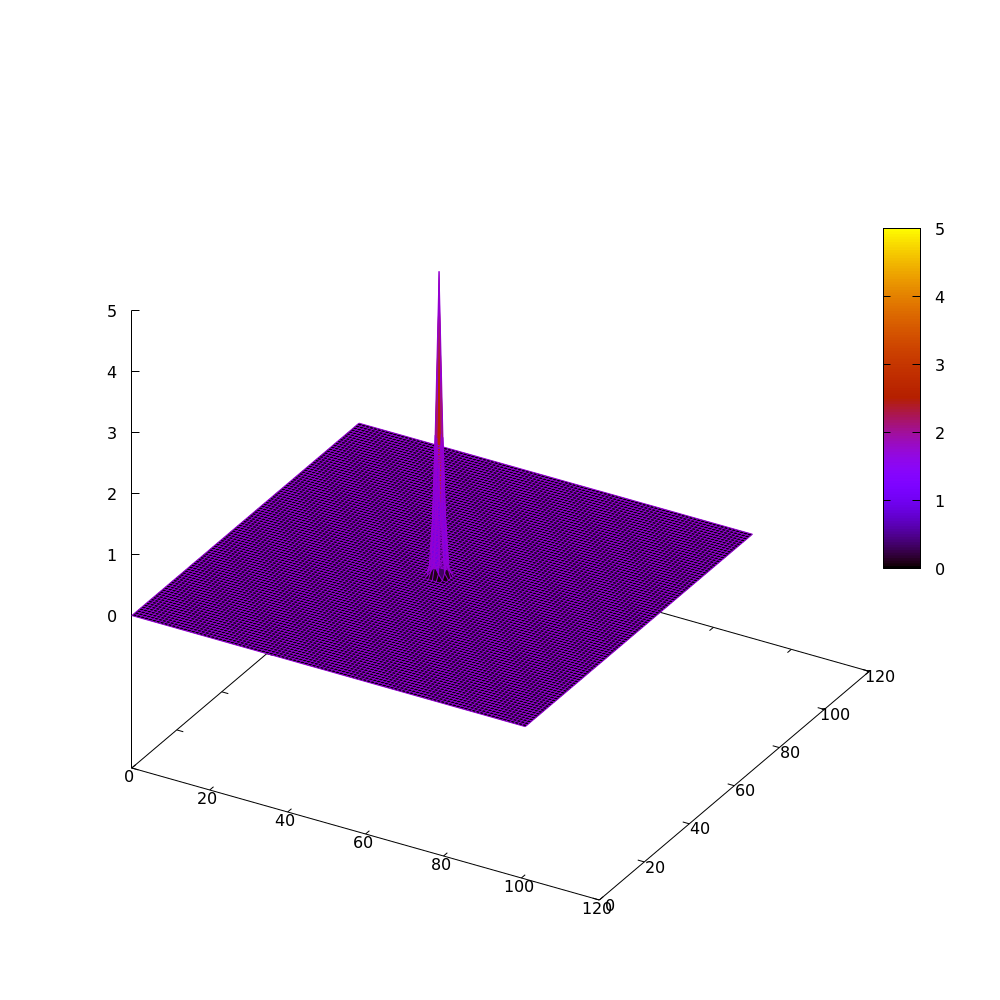}
                   
    \caption{This figure illustrates five Gaussians with different values of $\sigma$. It is known that when a sequence $\sigma_n$ converges towards zero, the corresponding Gaussian functions  converges towards the Dirac. From the physics point of view, the Dirac models a situation where the whole energy concentrates in a single point. The analogy with anger is the following: assume that the z-coordinate denotes the intensity of anger at each point of the psyche (represented by $\R^2$). At the beginning, the energy of anger is distributed and we do not feel it ($\sigma$ is large). But then, it starts to concentrate and the maximum level of energy increases and eventually leads to a pic of anger and a  loose of control (emotional putsch of anger).}
    \label{fig:Dirac-Anger}
\end{figure}

\section{The control of emotions under a mathematical perspective}
In this section, we would like to introduce a possible simple model for the control of emotions. Mathematically, the theory of optimal control  and the maximum principle of Pontryagin are the classical technique to tackle problems of control for which a time dynamic behavior is involved, see \cite{Book-Pont-1962,bookEvans2013,BookHocking1991}. Here, we describe a simple example, for which computations are explicit and which fits to the context of the control of anger. We also provide a solution of the problem considered.

Let's assume that following an external stimulus, the energy $E(x,y,t)$ of anger manifests in our conscious according to the following law:

\[E(x,y,t)=\dfrac{1}{\sigma(t)\sqrt{2\pi}}\exp{\bigg(-\dfrac{x^2+y^2}{2\sigma^2(t)}\bigg)}\]

with
\[\sigma_t=-(a-\alpha)\sigma\]
\[\sigma(0)=\sigma_0\]
\[\alpha(0)=0\]
Those  equations are set up in such a way that without control the energy will concentrate in a single point: the putsch of anger occurs and we loose control. We assume that the stimulus occurred at an initial time $t_i<0$, and that, at time $0$, we detect anger in a conscious way inside us and try to take over its manifestation. Here $a$ is a positive number and $\alpha(t)$ is a positive function which stands for the control one must to exert to control the manifestation of anger. Our admissible constraints read as 
\begin{equation}
\label{eq:E00t}
  E(0,0,t)=\dfrac{1}{\sigma(t)\sqrt{2\pi}}<E_M  
\end{equation}
where $E_M$ is the maximal intensity of anger acceptable. We look for the minimal $\alpha$ (the minimal effort)  such that at every time $t$ inequation \eqref{eq:E00t} is satisfied 
and such at the final time $T$ the energy 
\[E(0,0,T)=\dfrac{1}{\sigma(T)\sqrt{2\pi}}<E_s\]
where $E_s$ is an intensity which corresponds to a small common state of anger \textit{i.e.}, a state at which a person is calm. This means that at the final time $T$, one has succeeded to bring back the psychic state to a normal state. We denote 
\[\sigma_s=\dfrac{1}{\sqrt{2\pi}E_s},\]
\[\sigma_M=\dfrac{1}{\sqrt{2\pi}E_M},\]
and assume 
\[\sigma_M<\sigma_0<\sigma_s\]
meaning that the intensity of anger at time $0$, when we detect it, is less than the maximal intensity acceptable, but that the small common intensity of anger is less  than the initial intensity.    
We want to minimize the function
\[P(\alpha)=\int_0^T\alpha(t)dt\]
which stands for the total effort provided during the interval $[0,T]$, under the constraints
\[\sigma(T)\geq \sigma_s \mbox{ and for all }t \in [0,T], \sigma(t)\geq \sigma_M.\]
We denote this problem by $\mathcal{P}$.
Here, we allow $\alpha$ to take any value in $(0,+\infty)$ but will discuss this assumption below.
Under the above assumptions, the following theorem holds.
\begin{theorem}
Any non-negative function satisfying 
\begin{equation}
\int_0^T\alpha(t)dt=\ln\sigma_s-\ln\sigma_0 +aT   
\end{equation}
and  for all $t \in [0,T]$
\begin{equation}
\int_0^t\alpha(t)dt\geq \ln\sigma_M-\ln\sigma_0 +at   
\end{equation}
is a solution of the problem $\mathcal{P}$.
\end{theorem}
Furthermore, the following proposition holds.
\begin{Proposition}
Let 
\[c=\dfrac{\ln \sigma_s-\ln \sigma_0}{T}+a\]
and assume that
\[a^2-4c(\ln \sigma_0-\ln \sigma_M)<0\]
or 
\[a-\sqrt{a^2-4c(\ln \sigma_0-\ln \sigma_M)}>2Tc\]
then the function
\[\alpha^*(t)=ct\]
provides an optimal control for problem $\mathcal{P}$.
In this case, the optimal trajectory is given by
\[\sigma(t)=\sigma_0\exp(-at+0.5ct^2)\]
\end{Proposition}
The two above mathematical results provide therefore a solution to our problem of control of anger. However, note that in the above assumptions, $\alpha$ is allowed to take any value in $(0,+\infty)$, meaning that one may use an unlimited amount of energy to control its anger. One may also assume a limited energy for the control $\alpha$. For example, the following condition
\[\int_t^{t+h}\frac{\alpha(s)}{h}ds<C \mbox{ for all } t\in [0,T), \mbox{ and } h\in (0,T-t)\]
for some constant $C$ imposes a limited energy. 
In this case, the problem may have no solutions, for example for a large $a$. The meaning of that being that it is impossible for the person to take down its anger without any external additional energy. 
\begin{figure}
    \centering
    \includegraphics[width=4cm]{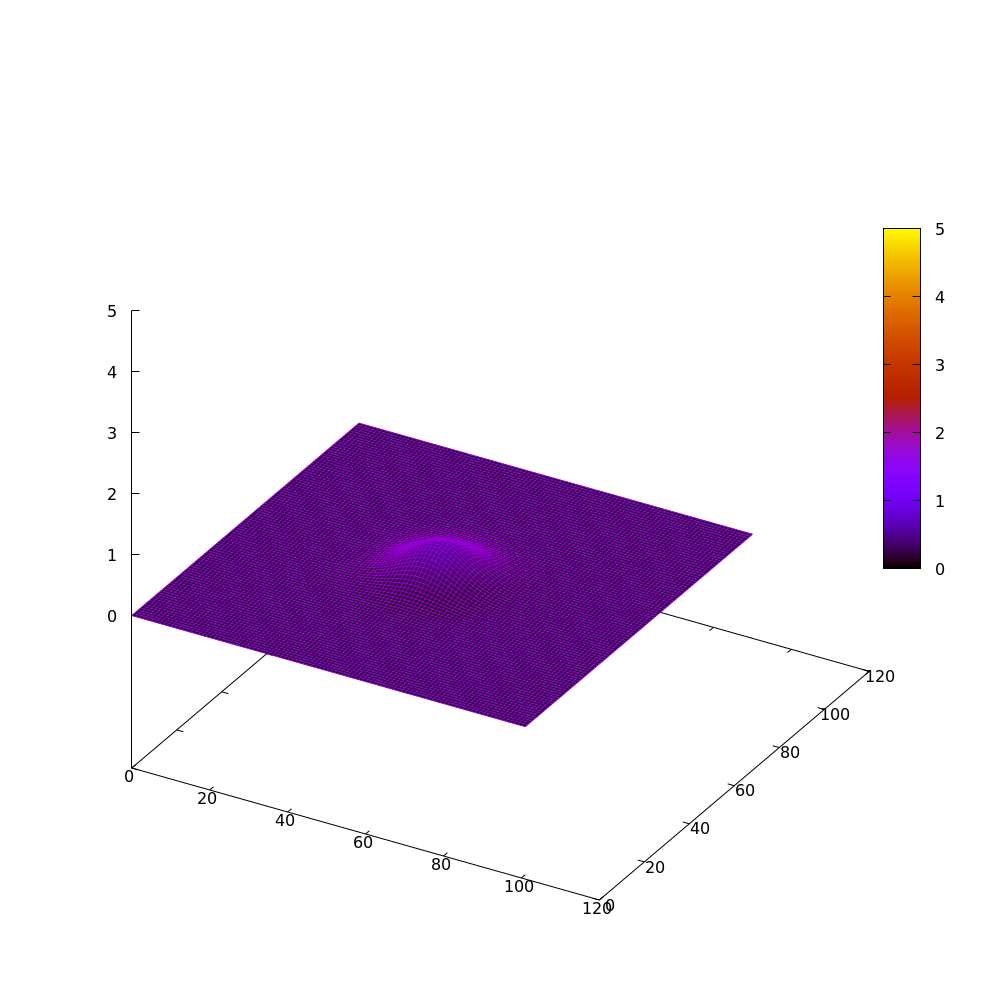}
       \includegraphics[width=4cm]{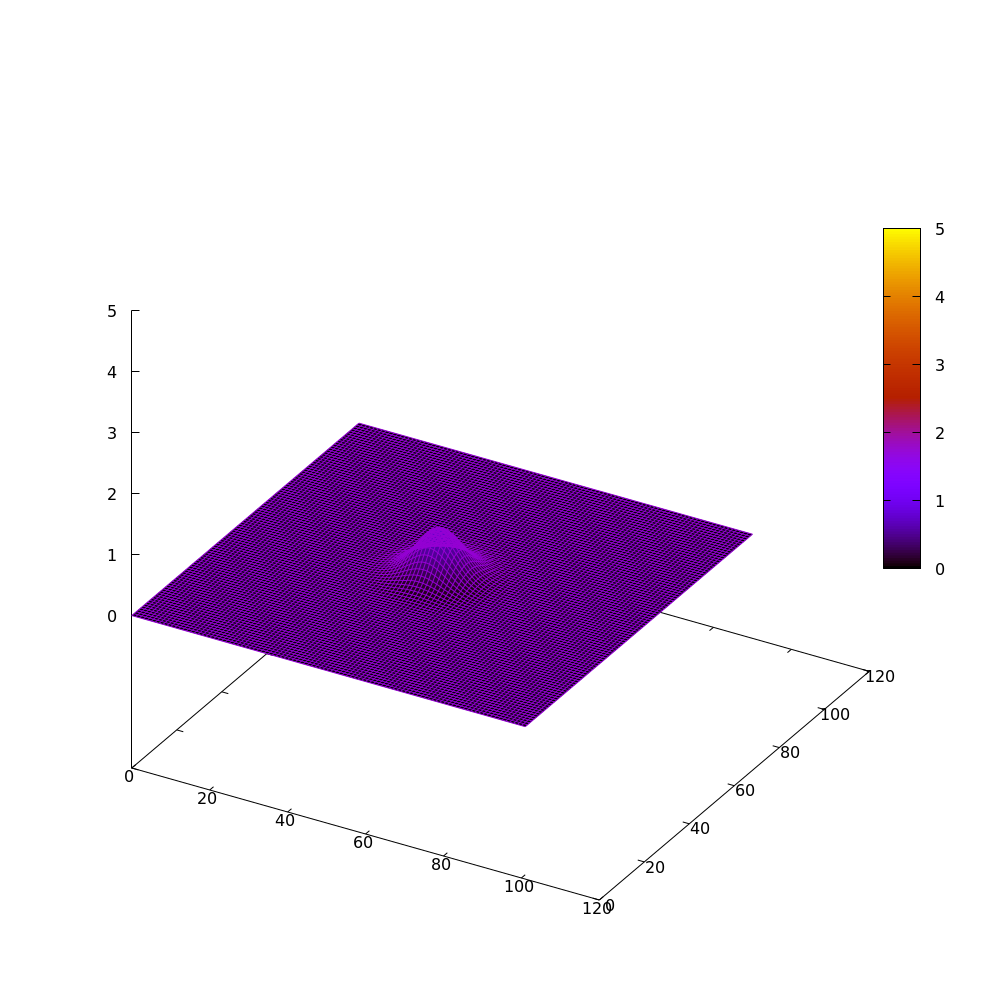}
          \includegraphics[width=4cm]{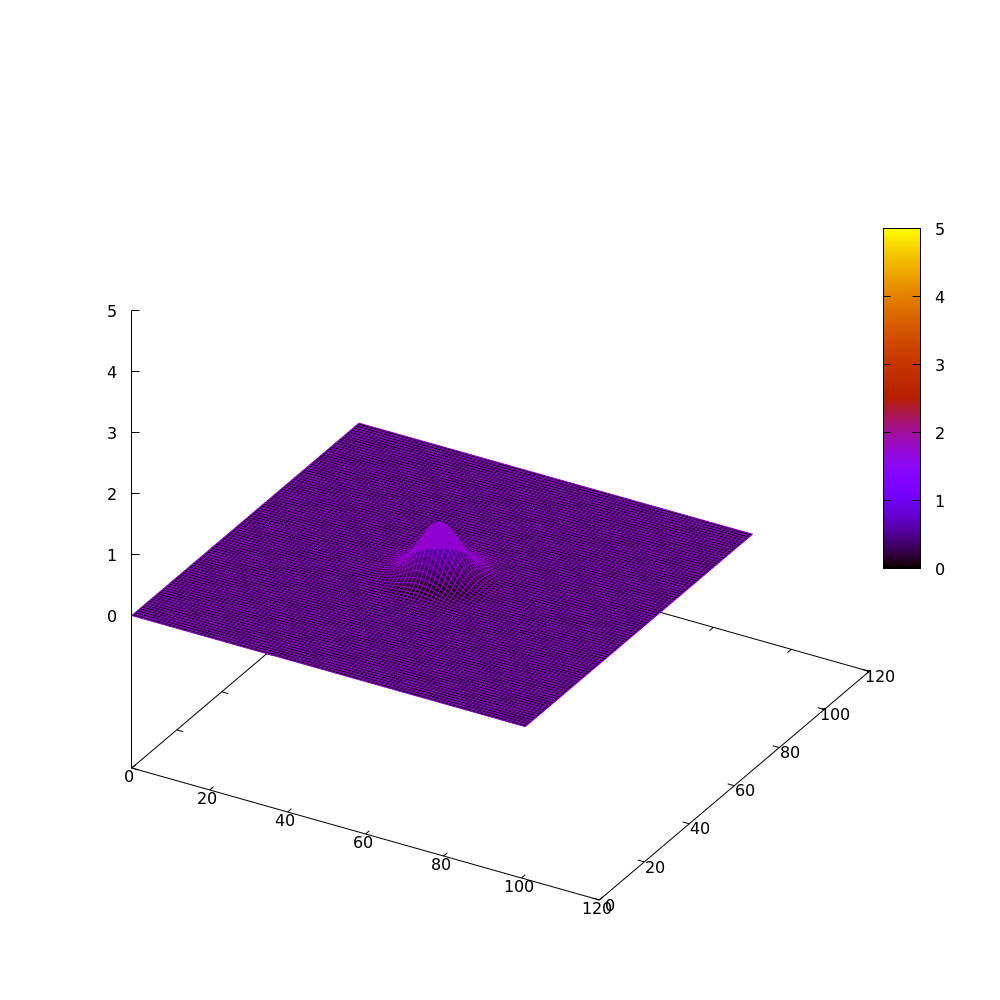}\\
             \includegraphics[width=4cm]{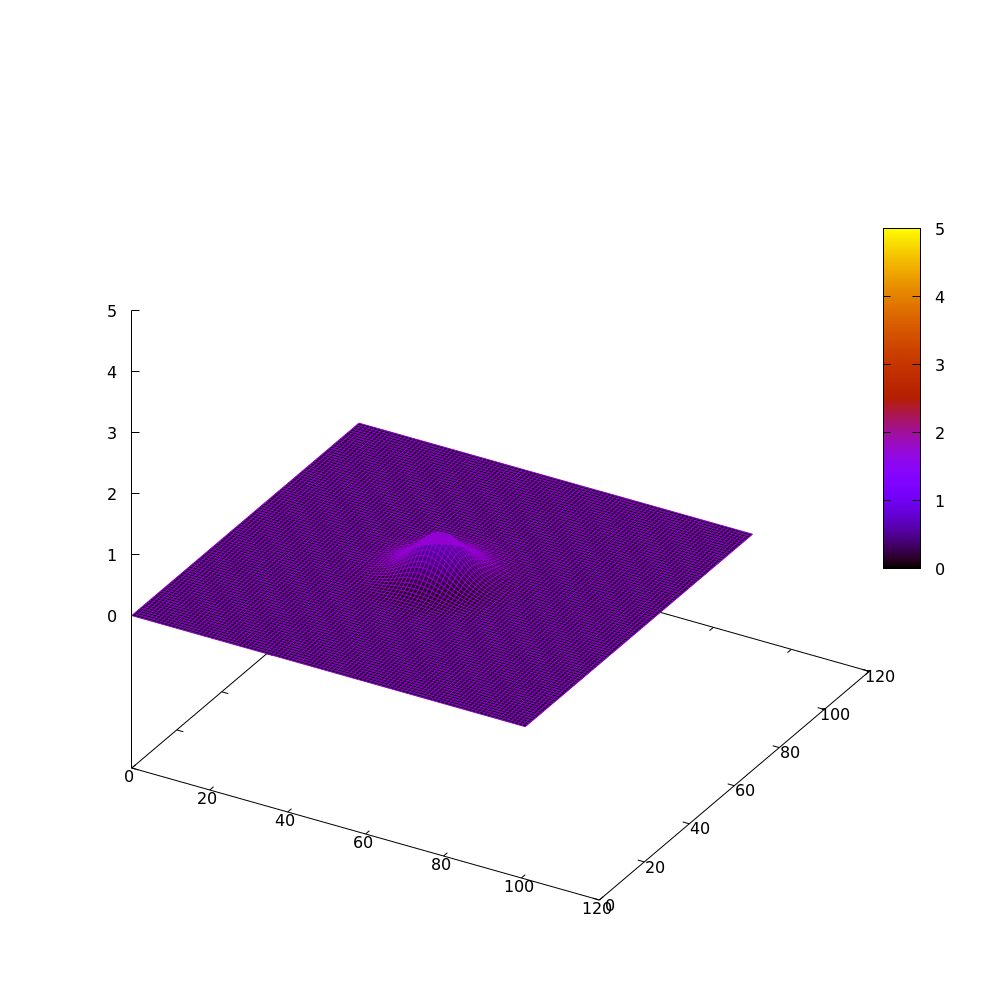}
                \includegraphics[width=4cm]{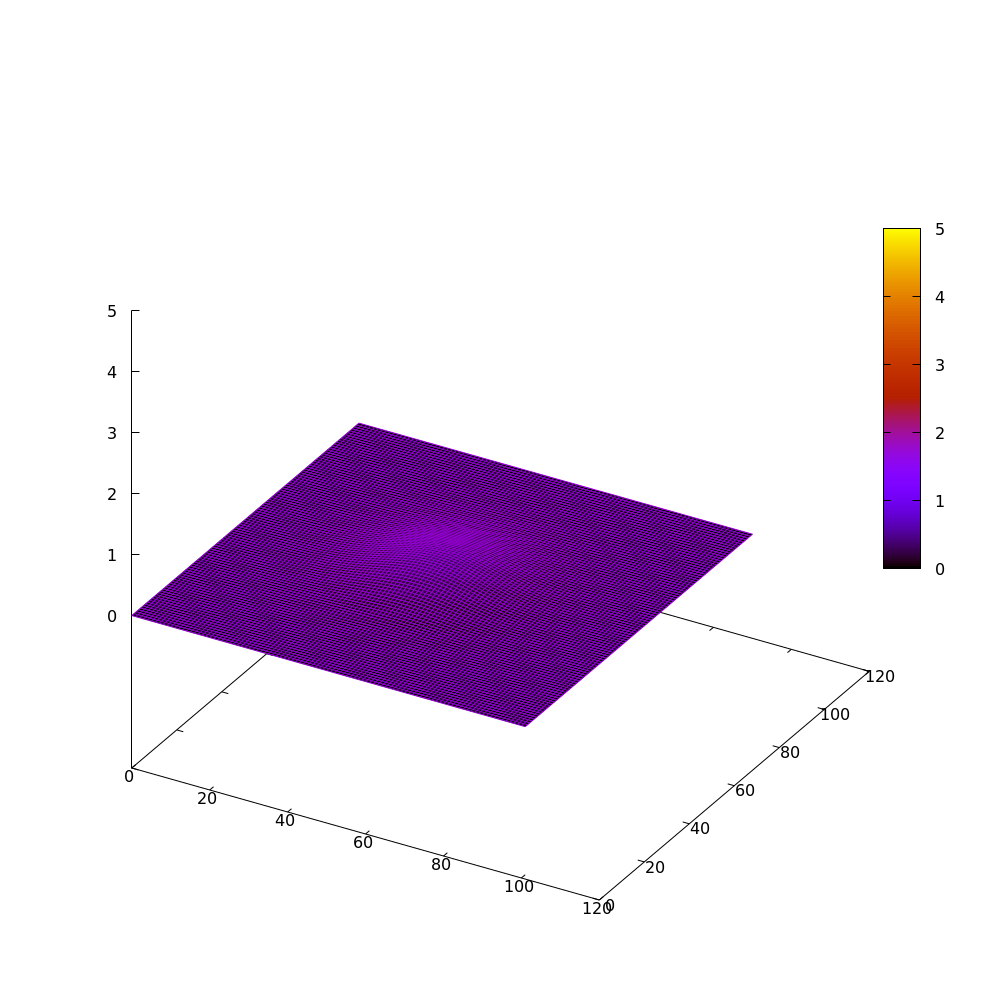}
                 \includegraphics[width=3cm]{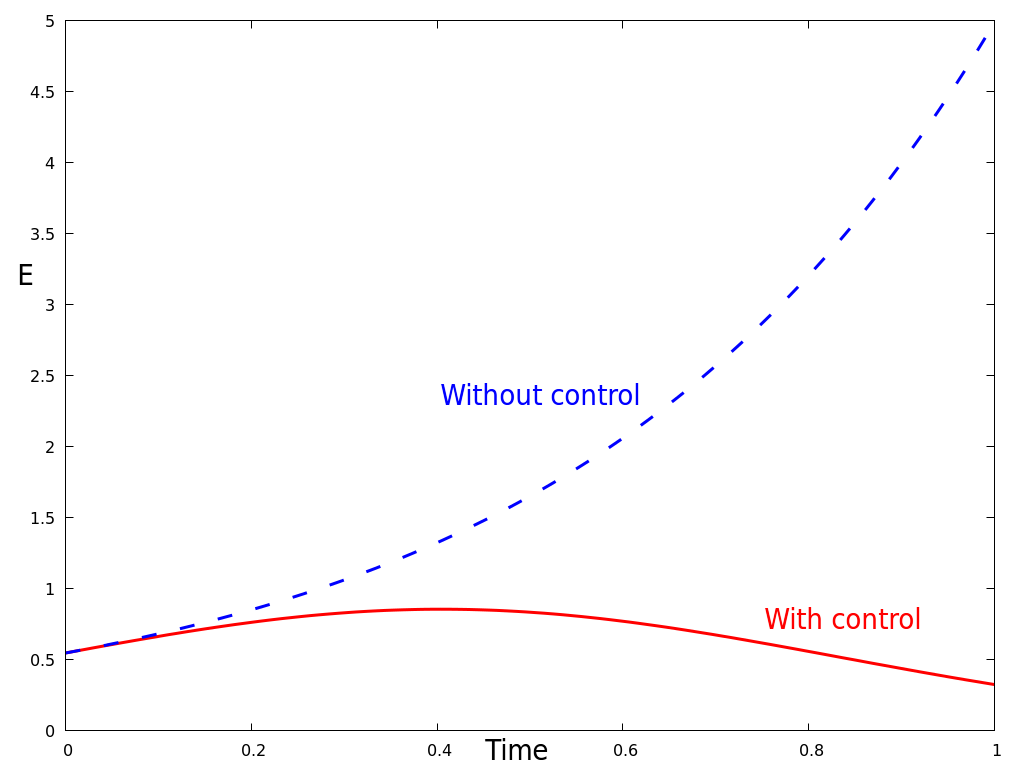}
                   
    \caption{This figure illustrates the control of a pulse of anger. Numerical illustration with $T=1, \sigma_i=3, \sigma_0=0.73, \sigma_s=1.23$ and a solution of the problem $\mathcal{P}$ given by $ \alpha^*(t)=ct$. Without control, the energy of anger would lead to a pic of anger and a loose of control. Applying the control $\alpha^*(t)$ allows to contain the energy of anger and to bring it to an acceptable level.}
    \label{fig:Control-Anger}
\end{figure}

\bibliographystyle{plain}
\bibliography{biblio}
\end{document}